\pdfoutput=1
\documentclass[%
reprint,
superscriptaddress,
amsmath,amssymb,
aps,
pra,
showkeys
]{revtex4-1}
\usepackage{graphicx}
\usepackage{dcolumn}
\usepackage{bm}
\usepackage{natbib}
\usepackage{hyperref}
\usepackage{float}
\usepackage{mathrsfs}
\usepackage{fancyhdr}
\usepackage{esint} % For 2D closure surface integration
\usepackage{accents} 
\usepackage{float}
\hypersetup{colorlinks=true, citecolor=blue, urlcolor=blue, linkcolor=blue}
\begin{document}
\lhead{}
\chead{}
\rhead{}
\lfoot{}
\cfoot{\thepage}
\rfoot{}

\title{Multiscale Modeling and Analysis for High-fidelity Interferometric Scattering Microscopy}

\author{Shupei Lin}
\affiliation{School of Physics and Wuhan National Laboratory for Optoelectronics, Huazhong University of Science and Technology, Luoyu Road 1037, Wuhan, 430074, People's Republic of China}

\author{Yong He}
\affiliation{School of Physics and Wuhan National Laboratory for Optoelectronics, Huazhong University of Science and Technology, Luoyu Road 1037, Wuhan, 430074, People's Republic of China}

\author{Hadrien Marc Louis Robert}
\affiliation{Institute of Photonics and Electronics, Czech Academy of Sciences, Chabersk\'a 57, 18251 Prague, Czech Republic}
%\affiliation{Institute of Photonics and Electronics, Czech Academy of Sciences, Chaberská 57, 18251 Prague, Czech Republic}

\author{Hong Li}
\affiliation{School of Physics and Wuhan National Laboratory for Optoelectronics, Huazhong University of Science and Technology, Luoyu Road 1037, Wuhan, 430074, People's Republic of China}

\author{Pu Zhang}
\affiliation{School of Physics and Wuhan National Laboratory for Optoelectronics, Huazhong University of Science and Technology, Luoyu Road 1037, Wuhan, 430074, People's Republic of China}

\author{Marek Piliarik}
\affiliation{Institute of Photonics and Electronics, Czech Academy of Sciences, Chabersk\'a 57, 18251 Prague, Czech Republic}

\author{Xue-Wen Chen}
\email[Corresponding author, Email: ]{xuewen\_chen@hust.edu.cn}
\affiliation{School of Physics and Wuhan National Laboratory for Optoelectronics, Huazhong University of Science and Technology, Luoyu Road 1037, Wuhan, 430074, People's Republic of China}

\date{\today}
\begin{abstract}
~\\
\noindent Interferometric scattering microscopy (iSCAT), as an ultrasensitive fluorescence-free imaging modality, has recently gain enormous attention and been rapidly developing from demonstration of principle to quantitative sensing. Here we report on a theoretical and experimental study for iSCAT with samples having structural dimensions that differ by 4-5 orders of magnitude. In particular, we demonstrate and intuitively explain the profound effects of sub-nanometer surface roughness of a glass coverslip and of a mica surface on the absolute signal and the shape of the point spread function of a gold nanoparticle. These quantities significantly affect the accuracies for determining the target size and position in all three dimensions. Moreover, we investigate a sample system mimicking a gold nanoparticle in a simplified cell environment and show position-dependent and even asymmetric point spread function of the nanoparticle. The multiscale study will facilitate the development of high fidelity iSCAT in real applications.
~\\

\end{abstract}

%\pacs{}
	
\keywords{Interferometric scattering microscopy, multiscale modeling, near-to-far field transformation, label-free, single-particle tracking}

\maketitle

%\centering	
%\includegraphics[width=7cm,height=3.5cm]{FigPDF/TOC}

\noindent 
Optical microscopy since its invention in seventeenth century has been an essential and ever-developing tool for visualizing matter noninvasively at the microscopic level in three dimensions \cite{mertz2019introduction, Weisenburger2015ConPhy}. Among various kinds of microscopy schemes \cite{mertz2019introduction, Weisenburger2015ConPhy}, interferometric scattering microscopy (iSCAT), relying on the interference of the scattering and a common-path reference light, has recently gained considerable interest as a ultrasensitive fluorescence-free imaging modality \cite{Lindfors2004PRL,Kukura2009NatMethods,Daaboul2010NanoLetter,Lee2018ACSPho,Yang2018PNSA,Holanova2019OLT,Ortega2012PCCP,hsieh2018OptComm,Young2019ARPC,Taylor2019NanoLetter}. Combining with effective background subtraction, iSCAT have demonstrated its superb detection sensitivity in a variety of contexts, for instance, directly detecting single viruses \cite{Kukura2009NatMethods,scherr2016ACSNano}, single exosomes \cite{Yang2018PNSA}, single molecules at room temperature \cite{celebrano2011NatPho} and single unlabeled proteins \cite{Piliarik2014NC,Ortega2014NanoLetter}, high-speed nanometer-precision single-particle tracking \cite{Hsieh2014JPCB,Spindler2016JPDAP}, quantitative mass measurement of single macromolecules \cite{young2018quantitative} characterizing the ultrafast carrier excitation and propagation in optoelectronic materials \cite{sung2020long}. As an important direction, iSCAT has been evolving from demonstrations of principle for ultrasensitivity towards quantitative sensing with high resolution and precision for practical applications, such as high-precision single-particle tracking in complex cell environment \cite{Taylor2019NatPho,Park2018ChemSci} and mass spectrometry of biological molecules \cite{Young2019ARPC}. The position and size (mass) of the nanoprobe are assessed through the iSCAT signal contrast and the shape of the point spread function of the nanoprobe in the sample system, which may involve materials with feature sizes differing by 4-5 orders of magnitude. In these circumstances, the acquired contrast and image patterns contain not only the signal from the nanoprobe but also contributions from other various parts, such as the substrate roughness of sub-nanometers, the cell membranes and a variety of cell interiors if the nanoprobe is in a cell environment. At first glance, the effect from the sub-nanometer surface roughness seems negligible. A closer study reveals a different view because the sub-nanometer height fluctuations occur unevenly, creating some domains with more hills than valleys or vice versa. The resulting hill domains, valley domains and the nanoprobe could have comparable scattering volumes but distinct scattering phases. Apparently, the various channels of contributions coherently will interact and change the signal contrast and the shape of the point spread function of the nanoprobe. Therefore, a rigorous multiscale analysis framework for iSCAT, which is currently lacking, become dispensable to facilitate the rapid development of iSCAT towards quantitative applications.
~\\
\begin{figure*}
	\centering
	\includegraphics[width=14.2cm]{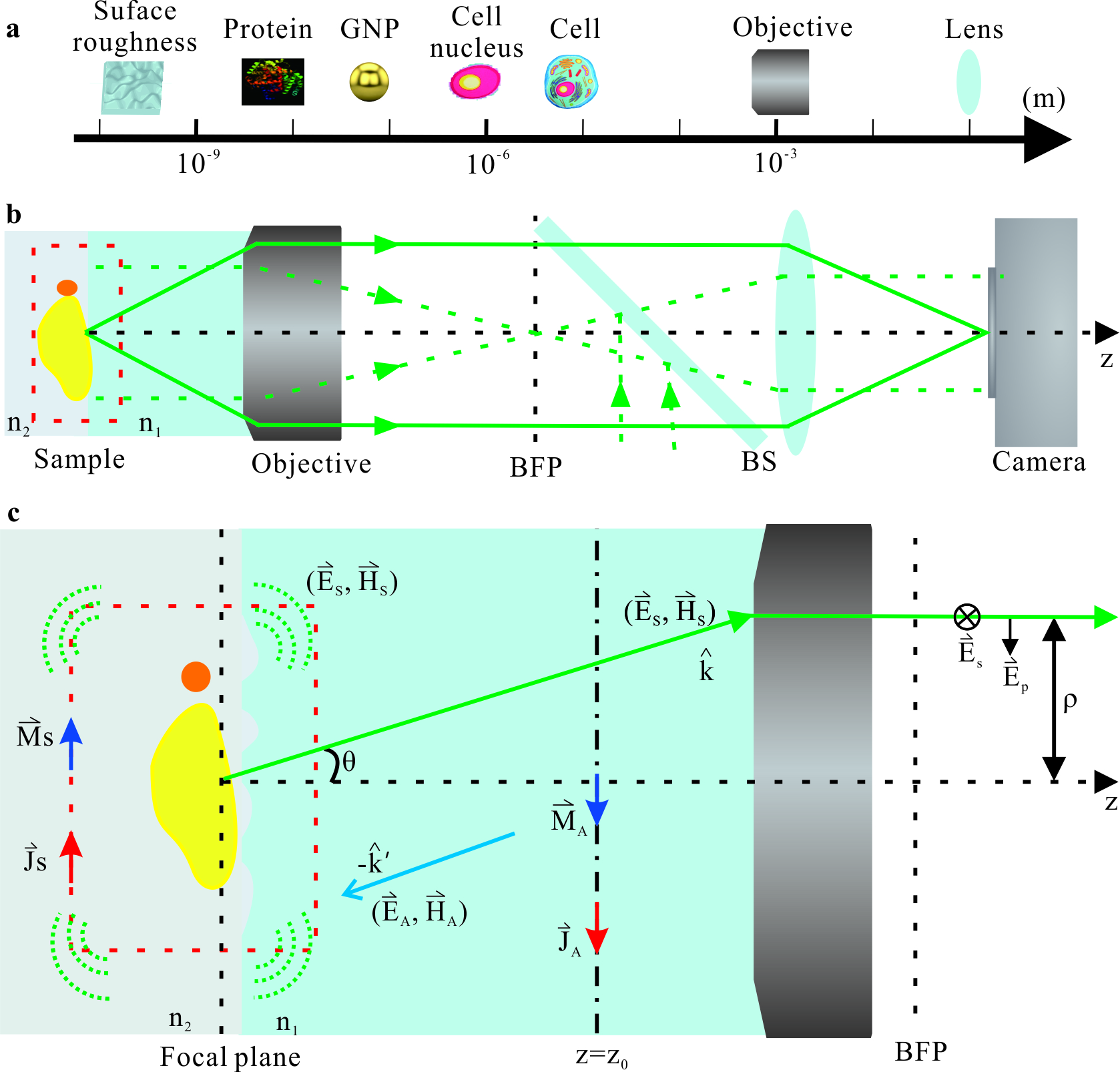}
	\caption{(a) Illustration of the multiscale scenario for iSCAT. (b) Schematic diagram of a reflective wide-field iSCAT setup. Back focal plane (BFP), Beam splitter (BS) (c) Illustrative sketch of the theoretical framework for the multiscale analysis (see text for details).}
	\label{Fig1}
\end{figure*}
~\\
\noindent Light propagation through an optical microscope system can be simulated by various approaches with different levels of accuracy, for instance, by ray-tracing methods \cite{torok1998general}, scalar diffraction and optical transfer-function approaches \cite{goodman2005introduction}, and vectorial diffraction imaging theory \cite{Wolf1959,torok2008OE,Foreman2011JMO,Totzeck2001Optik}. For modeling fluorescence microscopy with nanoscopic emitters such as single molecules, one could apply the Green’s tensor analysis and vectorial diffraction theory \cite{Backer2014JPCB,Novotny2012NanoOptics} and then retrieve the properties such as the dipole orientation \cite{bartko1999imaging,Lieb2004JOSAB} and position \cite{Kukura2009NatMethods,Zhang2013PC}. Naturally one may combine the above strategies together with the consideration of the illumination and reference fields to model iSCAT with single nanoparticles on ideally flat surfaces \cite{Avci2016OE,Trueb2017IEEEJSTQE}. However, such an approach falls short in dealing with iSCAT for sample systems that require multiscale modeling and analysis as previously discussed. 

~\\
\noindent 
Here we first present a general theoretical framework of multiscale modeling and analysis for iSCAT and then show the importance of the multiscale study for quantitative SCAT with concrete examples. In particular, for the first time, we demonstrate through rigorous simulations and experiments the profound effect of sub-nanometer surface roughness of a glass coverslip and of a mica surface on the interference contrast and shape of the point spread function of a single gold nanoparticle (GNP). Based on the theoretical formulation, we draw a transparent understanding of the observations why the domains of valley and bump of the roughness have different effects on the iSCAT image of the GNP. Moreover, we simulate a relatively large sample system mimicking a GNP in a simplified cell environment and show that how the interaction of the GNP with a dielectric nanoscale object may significantly modify its point spread function from symmetric to asymmetric. We discuss the implications of these findings for quantitative iSCAT applications. 

~\\
\noindent 
We begin our discussion with a schematic diagram in Figure \ref{Fig1}a illustrating the multiscale scenario of an iSCAT experiment. Surface roughness of a glass coverslip, small proteins, GNP labels, cell nucleus and cells have typical feature sizes of sub-nanometer, few nanometers, ~10 nanometers, ~100 nanometers and 1-10 micrometers, respectively. Figure \ref{Fig1}b shows a sketch of a simplified wide-field iSCAT setup, consisting of (from left to right) the sample, an objective, a beam splitter (BS), a lens and a camera for imaging. The excitation is launched from the bottom of the BS and focused to the back focal plane (BFP) of the objective for the wide-field illumination as indicated by the green-dashed traces. The scattering (green-solid traces) and the reflected illumination (green-dashed traces) are both collected by the objective and sent through the BS and the lens onto the camera, forming a common-path interference image. The essential part requiring multiscale analysis is the sample area indicated by the red-dashed rectangle and depicted with more details in \ref{Fig1}Figure c. 

~\\
\noindent \noindent The key task of the multiscale analysis is to obtain the angular spectrum of the electric field $A_i(\hat k)$ from the sample and to be collected by the objective, where $\hat k$is the unit wavevector related to the angle as shown in Figure \ref{Fig1}c and the subscript \textit{i} denotes the polarization state, which can be \textit{s} for \textit{s}-polarized or \textit{p} for \textit{p}-polarized planewaves, respectively. With the knowledge of $A_i(\hat k)$, the electric field propagating through the microscope imaging system can be traced and simulated by the well-established vectorial diffraction approaches \cite{Foreman2011JMO,Backer2014JPCB,Novotny2012NanoOptics}. To calculate $A_i(\hat k)$ due to samples with arbitrary shapes in a planar multilayer structure and any scheme of illumination, we apply the Lorentz reciprocity theorem\cite{Harrington2001,Zhang2019Nanoscale,yang2016near}. For clarity, as shown in Figure \ref{Fig1}c, we consider a half-space structure of two media ($\rm {n_2}$ and $\rm {n_1}$1) as the background structure and a sample system enclosed by a cuboid indicated with a red-dashed rectangle. The fields outside the cuboid can be considered to be generated by a set of surface (tangential) electric current $\accentset{\rightharpoonup}  J_{\rm S}$ and magnetic current $\accentset{\rightharpoonup}  M_{\rm S}$ on the six surfaces of the cuboid. The currents are related to the electromagnetic fields on the surface as $\accentset{\rightharpoonup}J_{\rm S} = \hat n \times \accentset{\rightharpoonup}H_{\rm S}$ and $\accentset{\rightharpoonup}M_{\rm S} = -\hat n \times \accentset{\rightharpoonup}  E_{\rm S}$, with $\hat n$ denoting the outward-pointing normal vector.

~\\
\noindent \noindent Note that here the incident fields (illumination) are excluded in $\accentset{\rightharpoonup}  E_{\rm S}$ and $\accentset{\rightharpoonup}  H_{\rm S}$ to retrieve the angular spectrum of the fields propagating towards to the objective. Then we introduce a set of auxiliary surface current $\accentset{\rightharpoonup}  J_{\rm A}$, $\accentset{\rightharpoonup}  M_{\rm A}$ at the far field in medium n1, which generate an incoming planewave in $-\hat{k}$ direction with a polarization state of i. Now according to the Lorentz reciprocity theorem \cite{Harrington2001} and the derivations given in Supporting Information \cite{SM}, the angular spectrum reads

\begin{equation}
A_i(\hat k)=-\frac{Z_1}{8\pi^2 \cos \theta}\varoiint_{\rm N}(\accentset{\rightharpoonup} H_{\rm S} \times \accentset{\rightharpoonup}  E_{\rm A}+\accentset{\rightharpoonup}  E_{\rm S}\times \accentset{\rightharpoonup}  H_{\rm A})\cdot \ d\accentset{\rightharpoonup}S
\label{Ai}
\end{equation}
where the enclosed integration is over all surfaces of the cuboid, $\theta$ is angle of the wave vector with respect to the $z$ axis (see Figure \ref{Fig1}c) and Z1 is the wave impedance of medium n1. $\accentset{\rightharpoonup}  E_{\rm A}$ and $\accentset{\rightharpoonup}  H_{\rm A}$ are the electric and magnetic fields at the surfaces of the cuboid due to the incoming planewave $-\hat{k}' $ (generated by the auxiliary surface currents) to the half-space structure and thus can be readily obtained in analytic form. $\accentset{\rightharpoonup}  E_{\rm S}$ and $\accentset{\rightharpoonup}  H_{\rm S}$ are numerically calculated by a finite-element method (FEM) with the application of a boundary conforming Delaunay triangulation for meshing in the platform of COMSOL Multiphysics \cite{SM}.
 
 ~\\
 \noindent \noindent
 The above paragraph describes how to calculate the angular spectrum of the electric field for arbitrary shapes of samples and any type of illumination schemes in a multilayered system. Next, we outline the steps to get the field expressions before the objective and at the imaging plane on the camera. We assume there is a small misalignment of the sample with respect to the focal point of the objective by $\Delta \accentset{\rightharpoonup}  d=(\Delta x,\Delta y,\Delta z)$. As shown in Figure \ref{Fig1}c, a spherical coordinate system with the focus of the objective as the origin is created and connected to a cylindrical coordinate system with optical axis of the objective as the rotation axis ($z$ axis). The electric field before the objective reads 
 \begin{widetext}
 	\begin{equation}
 	\accentset{\rightharpoonup} E_{\rm F}(r,\theta,\phi)=2\pi ik_{\rm{1z}}[A_s(\hat k_1)(\hat k_{\rm \rho}\times \hat z)+A_p(\hat k_1)(\hat k_{\rm \rho}\times \hat z)\times \hat k_1]\frac{e^{-ik_1r}}{r}e^{i\Delta \varphi}
 	\label{EF}
 	\end{equation}
 \end{widetext}
where $\Delta \varphi=k_1(\Delta x \sin \theta \cos \phi+\Delta y \sin \theta \sin \phi +\Delta z \cos \theta)$ is the phase shift of each planewave due to the misalignment. Following the vector ray-tracing and diffraction approach \cite{torok1998general,Backer2014JPCB}, the electric field at the back focal plane (BFP) of the objective $\accentset{\rightharpoonup}E_b (x,y)$ can be derived as shown in the Supporting Information \cite{SM}. Then the light intensity $I_m$ at the imaging plane at the position $(x_0, y_0)$ reads
\begin{equation}
\begin{aligned}
I_{\rm m}&=\left|\accentset{\rightharpoonup}E_{\rm m}(x_0,y_0)\right|^2
\\&=\left|\frac{i}{2\pi k_0f_0}\iint \,dx\,dy \accentset{\rightharpoonup}E_{\rm b}(x,y)e^{-i(\frac{k_0x_0}{f_0}x+\frac{k_0y_0}{f_0}y)}\right|^2
\end{aligned}
\label{Im}
\end{equation}
where the integration is over the BFP. Here $f_0$, and $k_0$ are the focal length of the lens before the camera and wavenumber in medium around the lens, respectively. By plugging the expression of $\accentset{\rightharpoonup}E_b (x,y)$ into Eq.\eqref{Im}, one can readily calculate the intensity distribution on the imaging plane and obtain the image of iSCAT contrast defined as ${I_{\rm m}}/{I_{\rm 0}}-1$, where $I_{\rm 0}$ is the intensity when there is no sample. With several lines of derivation \cite{SM}39, one could prove that the misalignment along the lateral direction leads to a shift of the image center by $\frac{n_1f_0}{n_0 f_1}(\Delta x,\Delta y)$, where $\frac{n_1f_0}{n_0 f_1}$ is the magnification factor with $f_1$ being the objective focal length. The longitudinal misalignment (defocusing)$\Delta z$ causes an angle (wavevector) dependent phase shift

\begin{equation}
\Delta \varphi=k_{\rm {1z}}\Delta z
\label{Delta_phi}
\end{equation}
where $k_1z$ is the axial component of the wavevector in medium $\rm{n_1}$. This explains the change of phase difference between the scattering and the reference light as a function of the defocusing parameter $\Delta z$. One may note that the changes of the interference contrast and pattern with respect to $\Delta z$ have been utilized to determine the longitudinal position of the nanoprobe \cite{Kukura2009NatMethods,hsieh2018OptComm,Young2019ARPC,krishnan2010Nat,mojarad2013OE}. 
\begin{figure}
	\centering
	\includegraphics[width=8cm]{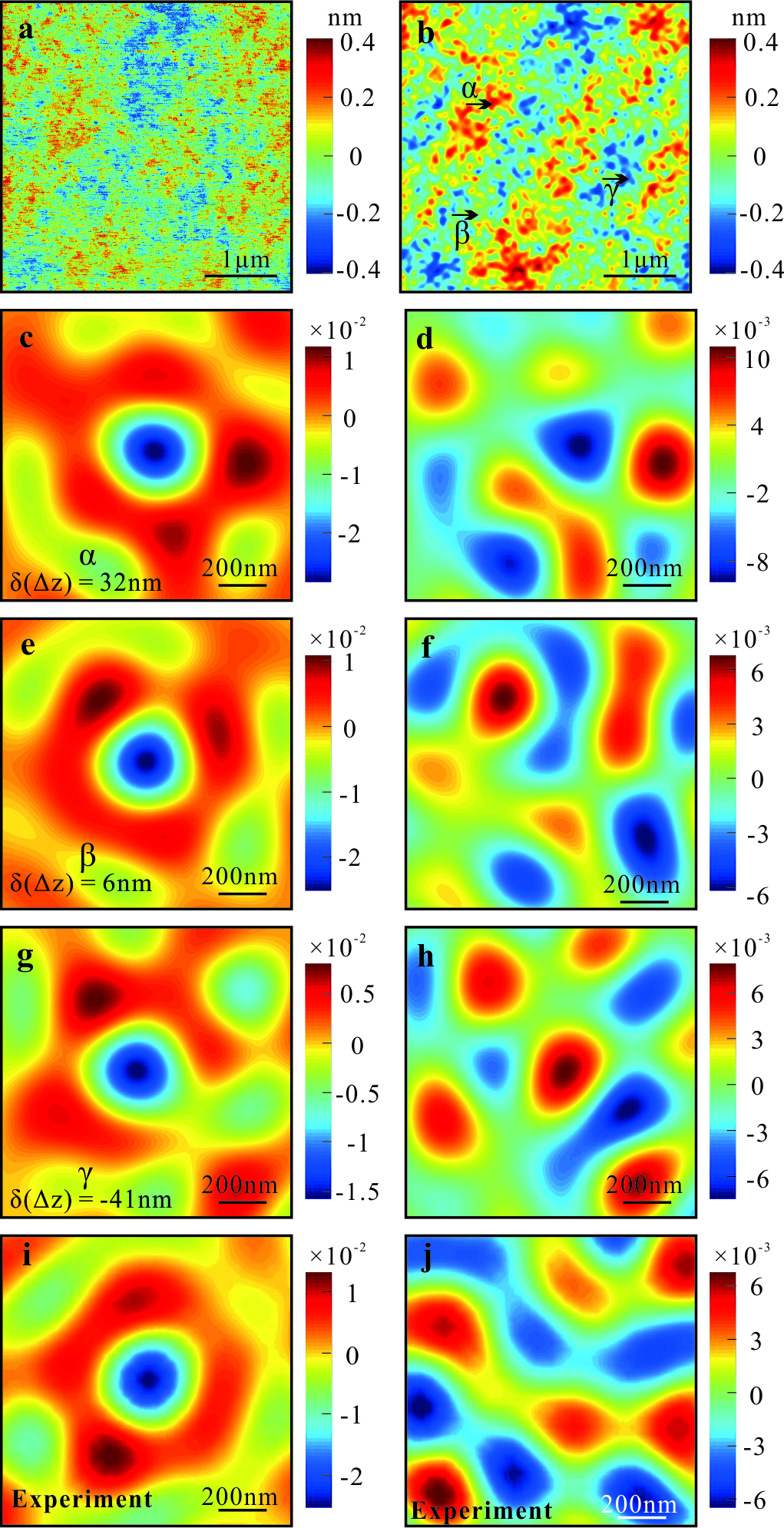}
	\caption{(a) Original AFM image from part of a glass coverslip surface. (b) Morphology image after low-pass filtering of the AFM image. $\alpha$, $\beta$ and $\gamma$ denote three locations with the arrows for iSCAT study. (c), (e) and (g) are the simulated iSCAT contrast images for a 20nm GNP on the coverslip at position $\alpha$, $\beta$ and $\gamma$, respectively. (d), (f) and (h) are the simulated iSCAT contrast images for the bare coverslip area centered around $\alpha$, $\beta$ and $\gamma$, respectively. (i) Measured iSCAT contrast image for the GNP on the coverslip. (j) Measured iSCAT contrast image for the part of bare coverslip area.}
	\label{Fig2}
\end{figure}

~\\
\noindent \noindent
In the next we apply the multiscale model to study the effect of sub-nanometer surface roughness of a glass coverslip and a mica surface on the iSCAT image of a single GNP. The first example concerns iSCAT with a 20nm GNP on a clean coverslip. Figure \ref{Fig2}a depicts an atomic force microscope (AFM) image for part of the coverslip. One observes that the height fluctuates within $\rm{\pm 0.4nm}$ and there exist domains of “valley” or “hill” across laterally several hundreds of nanometers. The more rapid height undulation observed in the AFM image (may be due to some electronic noises) makes little contribution to the scattering light since the scattering is proportional to the volume of the scatterer. To grasp the main contribution, we process the original AFM image with a low-pass spatial frequency filter \cite{SM} to obtain a smoother image as shown in Figure \ref{Fig2}b. In this way, the effect due to the surface roughness can be treated with an affordable computational demand. We consider three cases, \textit{i.e.}, the 20nm GNP positioned at a bump, a relatively flat area and a valley denoted by $\alpha$ $\beta$ and $\gamma$, respectively, as indicated in Figure \ref{Fig2}b. The iSCAT contrast images of the coverslip substrate with and without the 20nm GNP are calculated under normal wide-field incidence with a random polarization at the wavelength of 545nm. Figure \ref{Fig2}(c), \ref{Fig2}(e) and \ref{Fig2}(f) display the distorted iSCAT images by the roughness for the GNP placed at position $\alpha$, $\beta$ and $\gamma$, respectively. In these images, the defocusing parameter $\Delta z$ in each graph is optimized to provide the largest dip contrast at the center. We define the situation of the 20nm GNP on a perfectly flat glass surface as the reference case. We observe that changes of $\delta(\Delta z)$ with respect to the reference are 32nm, 6nm and -41nm for the GNP at position $\alpha$, $\beta$ and $\gamma$, respectively. Correspondingly, the central dip value changes from -2.2\% of the reference case to -2.8\%, -2.5\% and -1.6\%, respectively. Figure \ref{Fig2}(d), \ref{Fig2}(f) and \ref{Fig2}(h) show the speckle-like iSCAT images for the different parts of the bare glass coverslip centered around $\alpha$, $\beta$ and $\gamma$, respectively. The common feature is the existence of random domains of a few hundred nanometers with negative and positive contrasts, which will change sign as the defocusing parameter changes. Figure \ref{Fig2}(i) and \ref{Fig2}(g) depict the measured iSCAT images from experiments for the 20nm GNP on a coverslip and no particle on the coverslip, respectively. The contrast range and the image patterns for the two cases are quite similar to what one sees from the set of figures discussed above. In summary, one clearly observes that the glass surface roughness has a significant influence on the contrast image for the 20nm GNP such that one has to adjust the defocusing parameter at different locations of the surface to obtain the largest contrast. Moreover, the image pattern around the center become irregular with position dependent speckles, which may affect the accuracy of lateral localization.

\begin{figure}
	\centering
	\includegraphics[width=8cm]{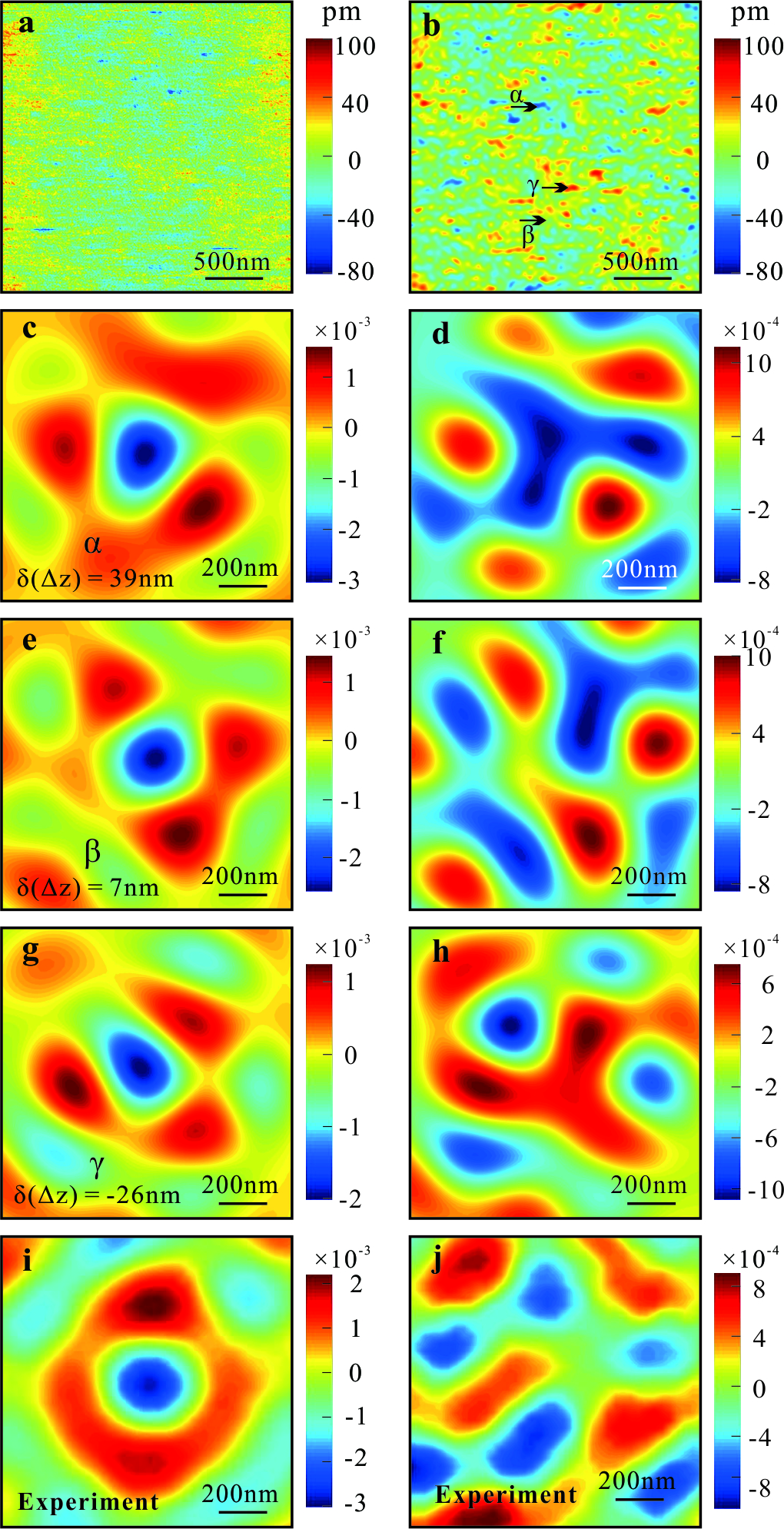}
	\caption{(a) Original AFM image from part of a mica surface. (b) Morphology image after low-pass filtering of the AFM image. $\alpha$, $\beta$ and $\gamma$ denote three locations with the arrows for iSCAT study. (c), (e) and (g) are the simulated iSCAT contrast images for a 10nm GNP on the mica at position $\alpha$, $\beta$ and $\gamma$, respectively. (d), (f) and (h) are the simulated iSCAT contrast images for the bare mica area centered around $\alpha$, $\beta$ and $\gamma$, respectively. (i) Measured iSCAT contrast image for the GNP on the mica. (j) Measured iSCAT contrast image for the part of bare mica area.}
	\label{Fig3}
\end{figure}

~\\
\noindent \noindent
Clearly, a smoother substrate will allow one to decipher smaller immobilized nanoparticles from the background via iSCAT, which motivates us to study both theoretically and experimentally the effectiveness of using an ultra-flat surface, for example, mica. By transferring a mica surface of $\sim$2 micron thick to a clean glass coverslip with index-matched oil, we are able to make a much smoother substrate surface. Figure \ref{Fig3}a shows an AFM image from the mica surface, which is indeed much smoother than the original coverslip surface. Nevertheless, there are still observable domains of “valley” and “hill” like the coverslip surface. Figure \ref{Fig3}(b) displays the surface morphology after low-pass filtering of the original AFM image and designates three different locations as before through $\alpha$, $\beta$ and $\gamma$, respectively. As shown in the Supporting Information \cite{SM}, the influence of the surface roughness of the mica surface to the iSCAT image of a 20nm GNP is much smaller than in the previous case. However, similar level of influence will be expected for a 10nm GNP. Figure \ref{Fig3}(c), \ref{Fig3}(e) and \ref{Fig3}(f) display the iSCAT images for the 10nm GNP placed at position $\alpha$, $\beta$ and $\gamma$, on the mica surface, respectively. Taking the 10nm GNP on a perfectly-flat mica surface as the reference case, we observe the largest interference dip changes from -0.24\% to -0.30\%, -0.26\% and -0.20\% for $\alpha$, $\beta$ and $\gamma$, respectively. Correspondingly, the changes of the defocusing parameter $\delta(\Delta z)$ with respect to the reference are 39nm, 7nm and -26nm, respectively. Figure \ref{Fig3}(d), \ref{Fig3}(f) and \ref{Fig3}(h) show the speckle-like iSCAT images for some area of the bare mica surface centered around $\alpha$, $\beta$ and $\gamma$, respectively. The patterns are similar to the glass coverslip case but with much smaller contrast modulations (about 8 fold smaller) due to the smoother surface. The experimental results are shown in Figure \ref{Fig3}(i) and \ref{Fig3}(j) for the cases of mica with a 10nm GNP and without a particle, respectively. One observes that the measured contrast ranges and the image patterns are very similar to the simulation results discussed above. 

\begin{figure}
	\centering
	\includegraphics[width=8cm]{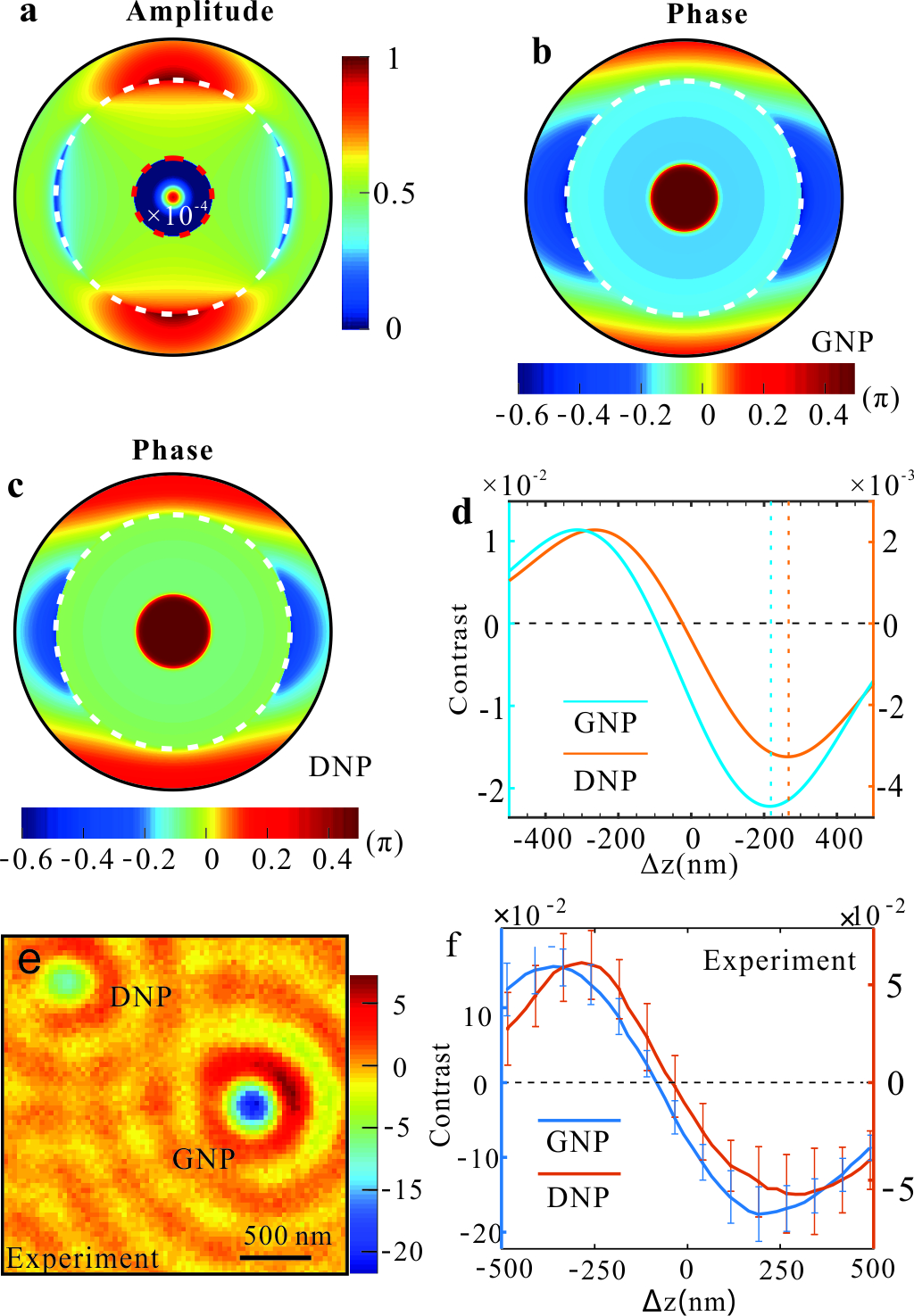}
	\caption{(a) Amplitude of the electric field at the BFP for a 20nm GNP on a perfectly-flat glass surface under wide-field illumination polarized along \textit{x} direction. Magnitude within the inner red-dashed circle is multiplied by $10^{-4}$. The white-dashed circle corresponds to NA = 1.0. (b) Phase of the electric field at the BFP for the GNP case. (c) Phase of the electric field at the BFP for the DNP case. (d) the iSCAT contrasts as functions of the defocusing parameter for the GNP and DNP cases. (e) Measured iSCAT contrast image of a 40nm GNP and a 50nm DNP immobilized on mica surface within one field of view. (f) Measured iSCAT contrasts as functions of the defocusing parameter for the 40nm GNP and 50nm DNP cases.}
	\label{Fig4}
\end{figure}

~\\
\noindent \noindent
Our theoretical model can draw a physical understanding of the phenomenon discussed in Figure \ref{Fig2} and \ref{Fig3}, such as the change of the interference dip value and the corresponding defocusing parameter for the GNP with the lateral location on the rough surface. We study two separate cases, \textit{i.e.}, a 20nm GNP and a 20nm dielectric nanoparticle (DNP) of silica on a perfectly-flat glass surface under a wide-field illumination with a polarization along $x$ direction and a wavelength of 545nm. Since iSCAT is a far-field microscopy scheme, the effect of a DNP is similar to an isolated nanoscale domain of “hill” of the roughness. Our model allows us to extract both the amplitude and phase of each planewave component of the light collected by the microscope objective. Figure \ref{Fig4}(a) display the amplitude distribution of the field at the BFP for the case of 20nm GNP on the glass surface. A bright spot within a red-dashed circle at the center (multiplied by $10^{-4}$) is due to the reference beam. The remaining part belongs to the scattering, corresponding to an in-plane dipole radiation above an interface \cite{Novotny2012NanoOptics}. From the simulation, one clearly observe that the reference beam and the scattering can be separated here and manipulated, for example, to increase the iSCAT constrast \cite{Cole2017ACSPho,Liebel2017NanoLetter,Avci2017Optica}. The amplitude pattern for the DNP case is almost identical \cite{SM}. However, the phase distributions at the BFP for the GNP case and DNP case, as shown in Figure \ref{Fig4}(b) and \ref{Fig4}(c) respectively, are quite different. Comparing these two graphs, one observes phase differences of about $0.1\pi$ for the scattering components, which originate from the difference of the dielectric constant of gold and silica. A defocusing of $\Delta z$ will introduce a phase shift for each planewave according to Eq.\ref{Delta_phi}. Thus the initial phase difference will lead to a shift of the interference contrast curve as a function of the defocusing parameter, which is illustrated clearly in Figure \ref{Fig4}(d) for the two cases. Compared to the GNP case, the interference dip of the DNP reaches maximum at a defocusing parameter of $\Delta z=264 \rm{nm}$, which is a shift of 48nm from $\Delta z=216 \rm{nm}$ of GNP. Similarly, we carried out the experiment with a 40 nm GNP and 50nm silica nanoparticle immobilized on the mica surface within one field of view as shown in Figure \ref{Fig4}(e). We extracted their contrast dependence on the defocusing $\Delta z$ and plot the dependence in Figure \ref{Fig4}(f) consistently with the model prediction. In the experiment we do not have the information about the absolute $\Delta z$ displacement. However, the relative shift between the interference dip of DNP and the GNP yields as much as (100$\pm$25) nm. Here the shift is larger than the simulation in Figure \ref{Fig4}(d) because in the experiment the particles are larger and the working wavelength is at 561nm \cite{SM}. Now considering the situation of a GNP on a bump like at position $ \alpha$, it is natural to have a positive shift of $\Delta z$ from the reference case to obtain a larger interference dip. The case of valley domain is similar to a negative scatter since a polarizability of scatter is proportional to the difference of the dielectric constant. Therefore, for the case of GNP on a valley domain, one should expect a negative shift of  $\Delta z$ from the reference case to obtain the best contrast. In addition, for this situation, the initial phase difference of two types of scattering is about 0.9$\pi$, thus one expects a reduced combined contrast. The conclusions drawn from the above analysis are the results we obtain consistently in Figure \ref{Fig2} and Figure \ref{Fig3}.

\begin{figure}
	\centering
	\includegraphics[width=8cm]{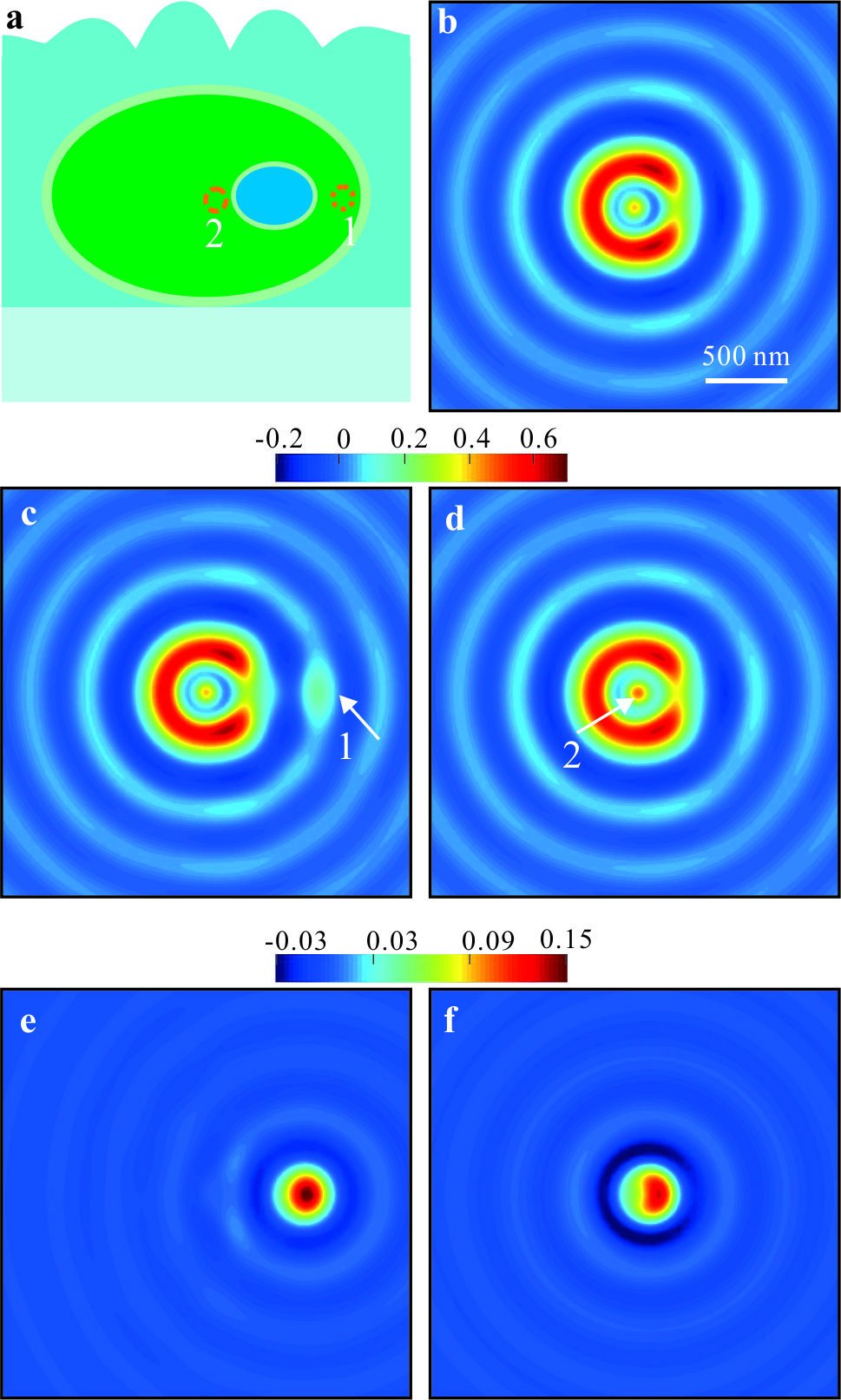}
	\caption{(a) Sketch of the sample structure. The 20nm GNP could be at position \#1 or \#2 as indicated by the red circle. Ellipsoids of the cell and cell nucleus have dimensions in the axes of (1.5$\mu$m,1.5$\mu$m,0.6$\mu$m) and (0.5$\mu$m,0.5$\mu$m,0.2$\mu$m), respectively. Simulated iSCAT contrast images for (b) the cell without the GNP, (d) the cell with GNP at position \#1, and (d) the cell with GNP at position \#2. (e) Contrast image due to the addition of GNP at position \#1. (f) Contrast image due to the addition of GNP at position \#2.}
	\label{Fig5}
\end{figure}

~\\
\noindent \noindent
The last case study tries to mimic the experiments using single GNPs as labels for tracking protein motions in a live cell \cite{Taylor2019NatPho}. For the sake of principle demonstration, we model the cell with a micron-size ellipsoid bounded by a 5nm-thick lipid membrane and assume the cell only has a nucleus with an ellipsoidal shape of smaller size bounded by another 5nm-thick lipid membrane and neglect other functional organelles. Figure \ref{Fig5}a shows the schematic diagram of the sample in water on a coverslip, where a 20nm GNP as the label may locate at different positions inside the cell. The refractive indices of the water, liquid inside the cell and nucleus, and membrane are 1.33, 1.36 and 1.46, respectively. In this example, we don’t take the effect of surface roughness into account since the scattering due to the roughness surface in an almost index-matched background is much smaller. Figure \ref{Fig5}b, \ref{Fig5}c and \ref{Fig5}d depict the contrast images for the cases without the GNP in the cell, with the GNP at position \#1 and with the GNP at position \#2. One observes that the contrast maps are mostly due to the strong signal from the cell and the GNP hardly can be directly identified. To get the GNP signal, we subtract Figure \ref{Fig5}c and Figure \ref{Fig5}d with Figure \ref{Fig5}b and obtain Figure \ref{Fig5}e and \ref{Fig5}f, respectively. Figure \ref{Fig5}e and \ref{Fig5}f display the point spread functions (PSFs) of the GNP at different locations relative to the cell nucleus inside the cell. The PSF is quite symmetric for position \#1 while it becomes asymmetric for position \#2, which is due to the stronger interaction with the cell nucleus at position \#2. One learns from these calculations that the PSF of small labels can be still retrieved by proper image processing although it may become irregular and strongly position dependent in the cell environment. The exact shape of PSF is important for high-precision single particle tracking and effective background subtraction in iSCAT \cite{Zhang2013PC,Cheng2017ACSPho}. 

~\\
\noindent \noindent
In summary, we have presented a holistic multiscale theoretical framework for modeling interferometric scattering microscopy with samples having structural dimensions different by up to 4-5 orders of magnitude. The modeling and analysis are based on rigorous electromagnetic numerical simulations, the Lorentz reciprocal theorem and vector-diffraction theory, and thus are applicable for any type of (structured) illumination and detection schemes for samples on a planar-multilayer substrate. The theoretical formulations allow transparent understanding of the optical image formed through the interference of the reference and scattering beams, including the origin of their phase-difference change with defocusing. The effects of substrate surface roughness for a normal glass coverslip and for a coverslip with a mica surface on single nanoparticle imaging have been rigorously modelled and compared with experimental observations for the first time. These studies demonstrate the significant influence of sub-nanometer surface roughness on the achievable signal contrast and the image pattern, particularly important for on-going efforts of using smaller and smaller labels or unlabeled nano-objects in practical applications. Moreover, we have rigorously simulated a relatively large system with small labels mimicking a gold nanoparticle in a micron-size cell, which may help to better understand the measured images and the behaviors of the point spread function in complicated environment like live cells. The numerical investigations of the point spread function in complex system will also be useful in developing deep-learning based rapid-background estimation by providing versatile training data data \cite{Zhang2018NatMethods,Mockl2020PNAS,Coker2019BioJ}. We believe the multiscale theoretical framework, rigorous modeling and analysis presented here will greatly facilitate the rapid development of interferometric scattering microscopy towards real applications.

\begin{acknowledgments}
\noindent We acknowledge financial support from the National Natural Science Foundation of China (Grant Number 11874166, 11604109), the Thousand-Young-Talent Program of China and the Ministry of Education Youth and Sports of the Czech Republic (project LL1602)
\end{acknowledgments}

\bibliographystyle{apsrev4-1}
\bibliography{Reference-Multi-Scale}
\end{document}